\def\doublespaced{\baselineskip=\normalbaselineskip\multiply
    \baselineskip by 175\divide\baselineskip by 100}
\def\tcg{$t$-$c$-$g$ }
\def\Ltcg{$\Lambda_{tcg}$ }
\def\sigtc{${\sigma_{tc}}^{max}$ }
\newcommand{\beq}{\begin{equation}}
\newcommand{\eeq}{\end{equation}}

\input{epsf.sty}
\documentstyle[12pt]{article}
\begin{document}
\begin{flushright}
MSUHEP - 61105 \\
\end{flushright}
\vspace{3 mm}
\large
\begin{center}
{Anomalous \tcg coupling: The Connection between Single Top
Production and Top Decay}\\
\vspace{5 mm}
\large
 {Tim Tait and C.--P. Yuan}\\
\small
\vspace{3 mm}
Department of Physics and Astronomy,
Michigan State University \\
East Lansing, MI 48824, USA\\
\vspace{5 mm}
\end{center}
\begin{abstract}
Continuing earlier work, we examine the constraint on an anomalous
\tcg coupling from top quark decay.  We find that from current
CDF measurements of the branching ratio $t \rightarrow W b$,
the minimum scale at which new physics can strongly modify the
\tcg coupling is \Ltcg $\geq$ about 950 GeV.  At the upgraded
Tevatron, single top production can constrain \Ltcg $\geq$ 4.5 TeV.
The connection between $t$-$c$
production and the decay $t \rightarrow c g$ is examined, showing
how constraints on one lead to a constraint on the other.
 \end{abstract}
\normalsize
\doublespaced

As shown in \cite{us}, an anomalous coupling between top quark,
charm quark, and gluon fields can have a large effect on single top
production at the Tevatron\footnote{ The notation
and conventions used in this work are the same as those presented
in \cite{us}.  A full discussion of the effective operator which
generates the \tcg coupling can be found there-in.}.
In that work, it was mentioned that the
\tcg operator can also be studied from the decay $t \rightarrow c g$.
This decay mode has been
studied in \cite{them}, and the results are similar to those presented
here.

In \cite{us}  the ratio of the partial widths is presented,
\beq
R_{tcg} = \frac{\Gamma( t \rightarrow c g)}{\Gamma (t \rightarrow W^{+} b)}
=\frac{\sqrt{2} \;64 \alpha_{s} \pi m_{t}^{2}}{3 \Lambda^{2} G_{F}
\left[1 - \frac{m_{W}^{2}}{m_{t}^{2}} \right] ^{2}
\left[1 + 2 \frac{m_{W}^{2}}{m_{t}^{2}}\right] }.
\eeq
Assuming that there is no other new physics modifying the top
quark decays, the branching ratio BR($t \rightarrow W b$) can be
expressed as,
\beq
{\rm BR}( t \rightarrow W b) = \frac{\Gamma (t \rightarrow W b)}
{\Gamma (t \rightarrow W b) + \Gamma (t \rightarrow c g)}.
\eeq
From these results, one may determine the \Ltcg which corresponds
to a given lower limit on BR($t \rightarrow W b$), $b_{min}$.  The
results are plotted in Figure-1.

\vspace{6 mm}
\centerline{\hbox{
\epsfysize=60 mm \epsfbox{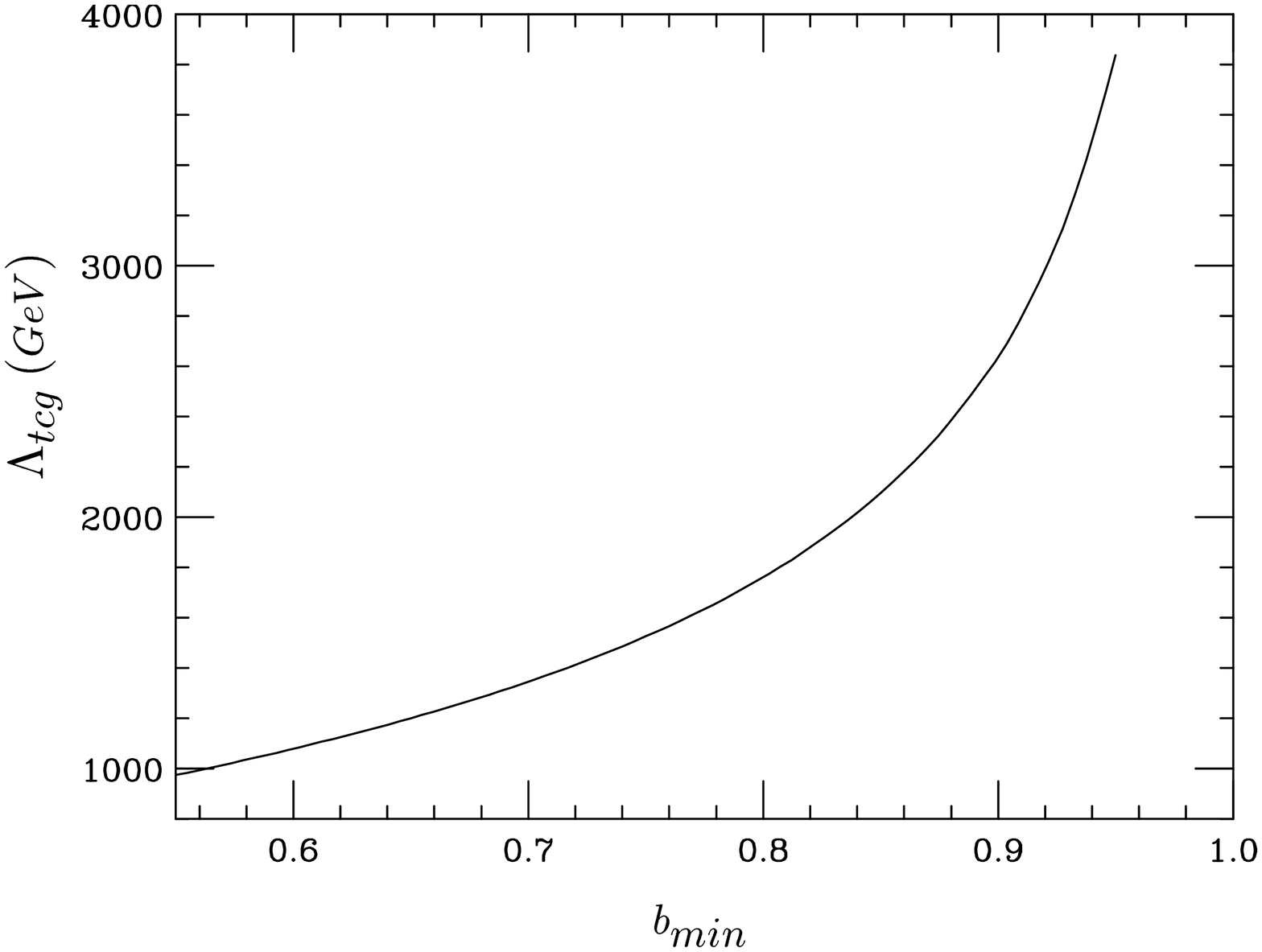}} }
\label{fig1}

Figure 1: The constraint on \Ltcg as a function of $b_{min} $.
\\

The current CDF measurement of BR($t \rightarrow W b$),
\beq
{\rm BR}( t \rightarrow W b) =
0.87^{\pm0.13\pm0.13}_{\pm0.30\pm0.11}
\eeq
leads to the 1 $\sigma$ limit $b_{min}$ = 0.55.  From this we derive
a constraint on \Ltcg $\geq$ 950 GeV.  As shown in \cite{us},
the maximum effect to the top-charm production rate (99\% C.L.) is,
\beq
{\sigma_{tc}}^{max} = 616 \;\left( \frac{ {\rm TeV}}
{\Lambda_{tcg}} \right)^2\; {\rm fb},
\eeq

including some basic acceptance cuts.  From the constraint on \Ltcg
discussed above, this implies \sigtc $\leq$ about 0.6 pb.  The
relation between $b_{min}$ and \sigtc is plotted
in Figure-2.  At the upgraded Tevatron (Run-II), the integrated luminosity
is expected to be 2 ${\rm fb}^{-1}$ per year.  As shown in \cite{us}, if no
$t$-$c$ signal events are observed
in 2 ${\rm fb}^{-1}$ of integrated luminosity,
this would imply \Ltcg $\geq$ 4.5
TeV, and from Figure-1, this requires $b_{min} =$ 96.3\%
(BR($t \rightarrow c g$) $\leq$ 3.7\%).

\vspace{6 mm}
\centerline{\hbox{
\epsfysize=60 mm \epsfbox{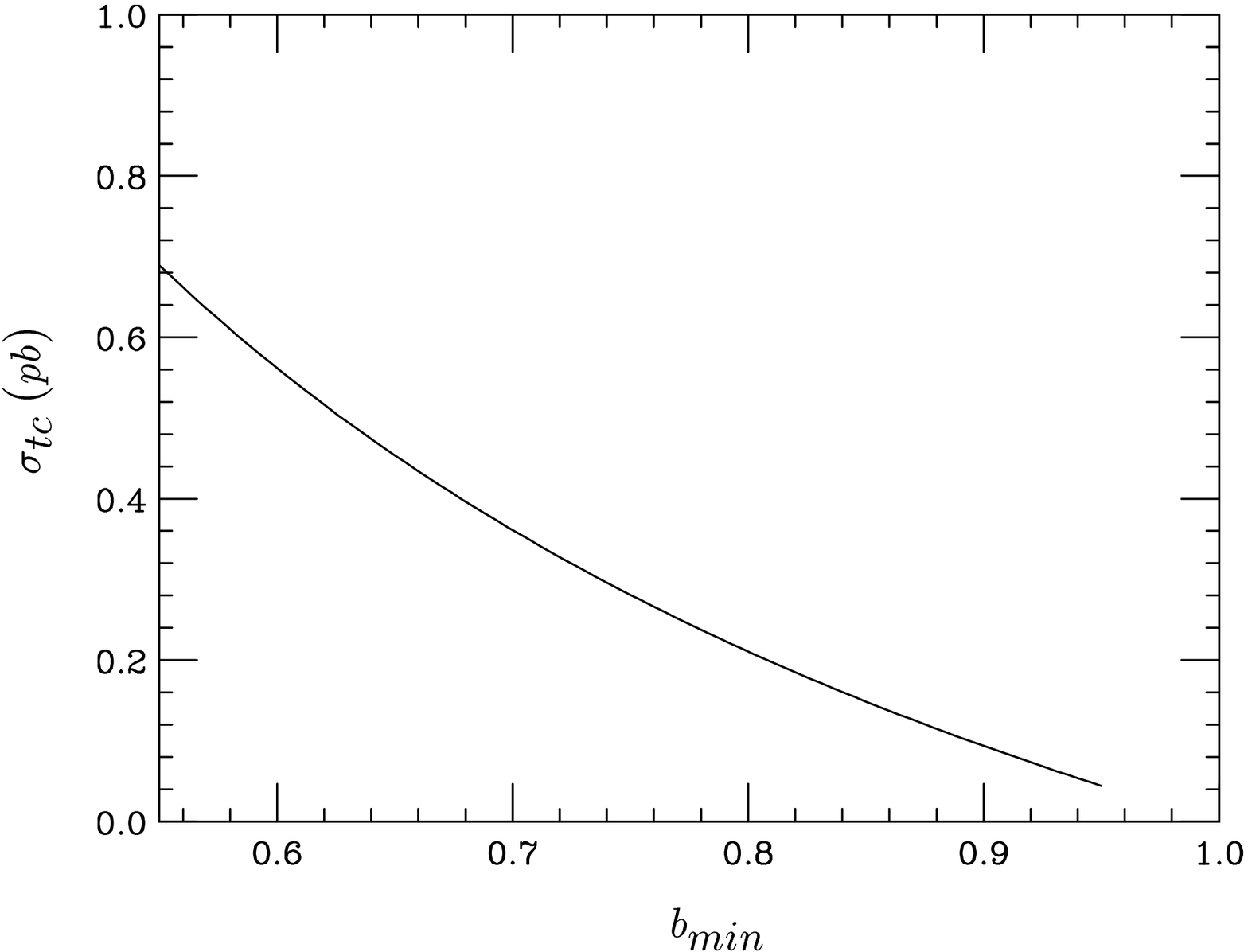}} }
\label{fig2}
Figure 2: The maximum cross section, \sigtc as a function of $b_{min}$ .
\\

The connection between \sigtc and $b_{min}$ is interesting,
because it gives a prediction of one quantity in terms of the other one.
This can serve to help identify this type of new physics effect from
other possible effects, and shows how a constraint derived from one
process has implications for another one.

Recently detailed work has been done on studying the measurement of
\Ltcg from top decays, including
kinematic cuts to optimize the signal to background
ratio \cite{them2}.  These results show studying top decays to be a
promising way to constrain \Ltcg, almost as promising as studying
single-top production for the current data sample, and possibly
providing even stronger constraints at the upgraded Tevatron.

The authors would like to thank  Ehab Malkawi and X. Zhang
for helpful discussion.
This work was supported in part by the NSF
under grant numbers PHY-9309902 and PHY-9507683.

\end{document}